\documentclass[11pt]{article}
\usepackage{wds11,epsf,multirow}

\usepackage[square]{natbib}

\lefthead{TORBANIUK ET AL.}
\righthead{DEPENDENCE BETWEEN SOME QUASAR CHARACTERISTICS}

\begin{document}

\title{Dependence between some spectral and physical characteristics of quasars}

\author{O. Torbaniuk$^{1,2}$ \& G. Ivashchenko$^{2,3}$}

\affil{$^{1}$Faculty of Physics, Taras Shevchenko National University of Kyiv, Kyiv, Ukraine \\
$^{2}$Main Astronomical Observatory of the National Academy of Science of Ukraine, Kyiv, Ukraine \\
$^{3}$Astronomical Observatory, Taras Shevchenko National University of Kyiv, Kyiv, Ukraine.}

\begin{abstract}
Using 192 composite spectra stacked from subsamples of individual SDSS DR7 quasar spectra binned in spectral index, $\alpha_{\lambda}$, and logarithm of monochromatic luminosity at 1450\,\AA, $\log{l_{1450}}$, we found that there is a dependence of the emission line equivalent width (EW) on spectral index (correlation or anti-correlation) for some lines, mostly for those lines for which the Baldwin effect is detected. We also show that there is no dependence between the virial mass of the central supermassive black hole of a quasar and its spectral index $\alpha_{\lambda}$.
\end{abstract}

\begin{article}

\section{Introduction}

The ultraviolet-optical spectral energy distribution (SED) of quasars is usually called the ``Big Blue Bump'' and characterised by a smooth continuum, which is considered to be thermal emission from an accretion disk, emission lines attributed to this emission reprocessed in surrounding clumped gas in a form of broad and narrow line regions as well as in a dust obscured torus, broad absorption lines (possessed by about 10\% of quasars) related to some streams outward quasar centre, and sets of narrow absorption lines caused by intergalactic medium. Despite of general similarity in UV-optical SED of quasars, they differ in equivalent width (EW) of emission lines and spectral index $\alpha_{\lambda}$. The proximity of the regions producing emission in continuum and absorption lines (i.\,e. the torus and the clumped broad line region, BLR) is considered to be the most promising explanation of the \cite{baldwin+77} effect: the inverse correlation of EW of some emission lines with the monochromatic luminosities at UV region. This effect was found, e.\,g., for  Ly$\alpha$ (1215~\AA), Si{\sc ii} (1260~\AA),  O{\sc i} (1304~\AA), Si{\sc iv}+O{\sc iv}] (1400~\AA) lines (see e.\,g. \cite{dietrich+02,green+01,shang+03}). But the physical explanation of the difference in $\alpha_{\lambda}$ and relation of quasar luminosity and lines' EW to $\alpha_{\lambda}$ are still not clear. 

Previously, using composite spectra of quasars with different $\alpha_{\lambda}$, we showed that there is no correlation between $\alpha_{\lambda}$ and monochromatic luminosity at 1450\,\AA, $\log{l_{1450}}$ (\cite{ivashchenko+13}). We also have not found any dependence of EW of lines within 1210--1450\,\AA\ on spectral index $\alpha_{\lambda}$ (calculated within the range 1270--1480~\AA) (\cite{torbaniuk+12}). But due to the fact, that the latter study was carried out without separation by luminosity, its result requires verification on a sample of spectra with different luminosity as well.  Such verification has been done in the present study. We also have checked the possible relation between the spectral index and the mass of the central black hole. 

\section{The data}
\begin{figure}[!h]
 \begin{center}
 \begin{tabular}{c}
   \epsfxsize=77mm
  \epsfbox{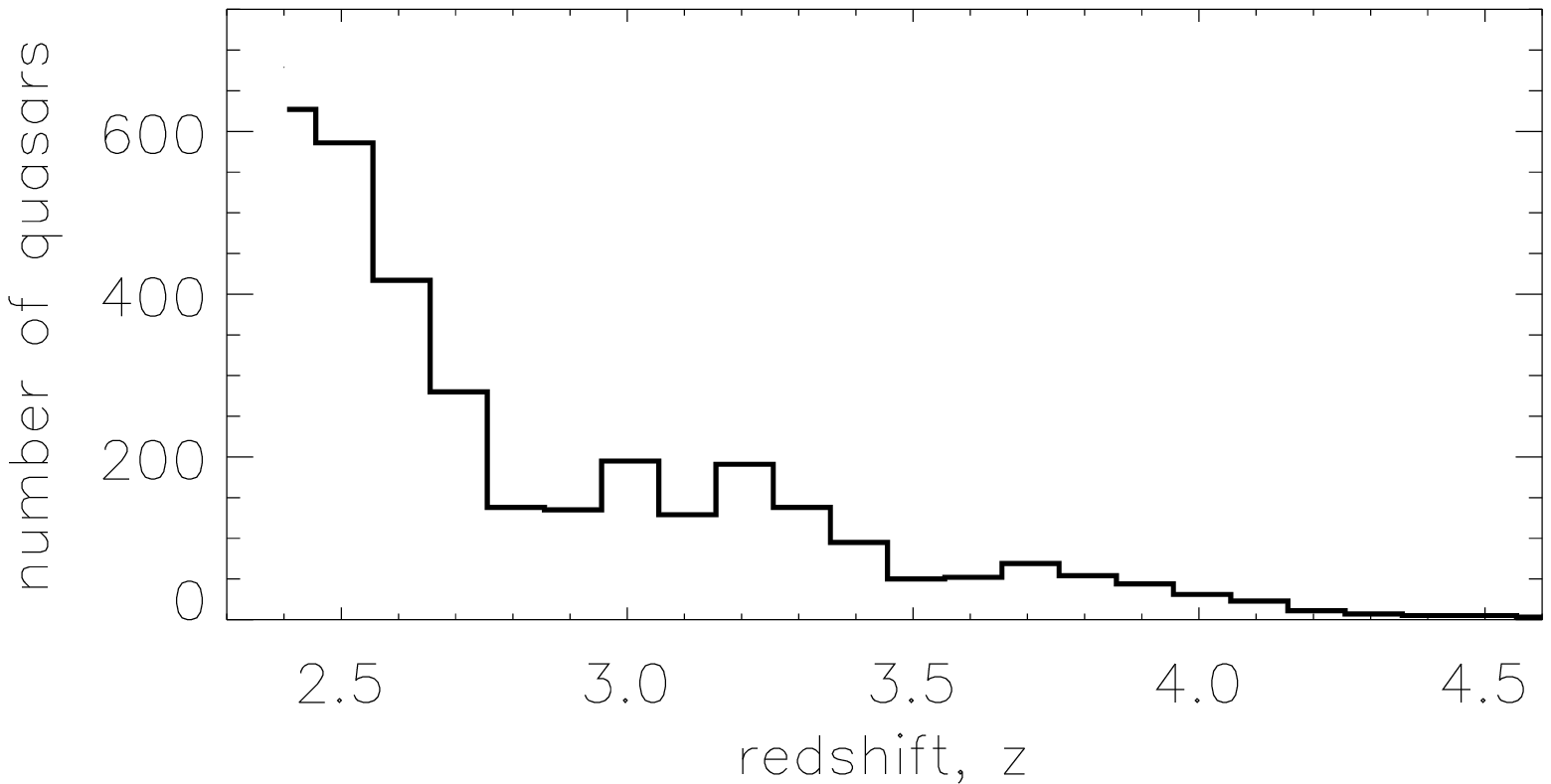}
   \epsfxsize=77mm
\epsfbox{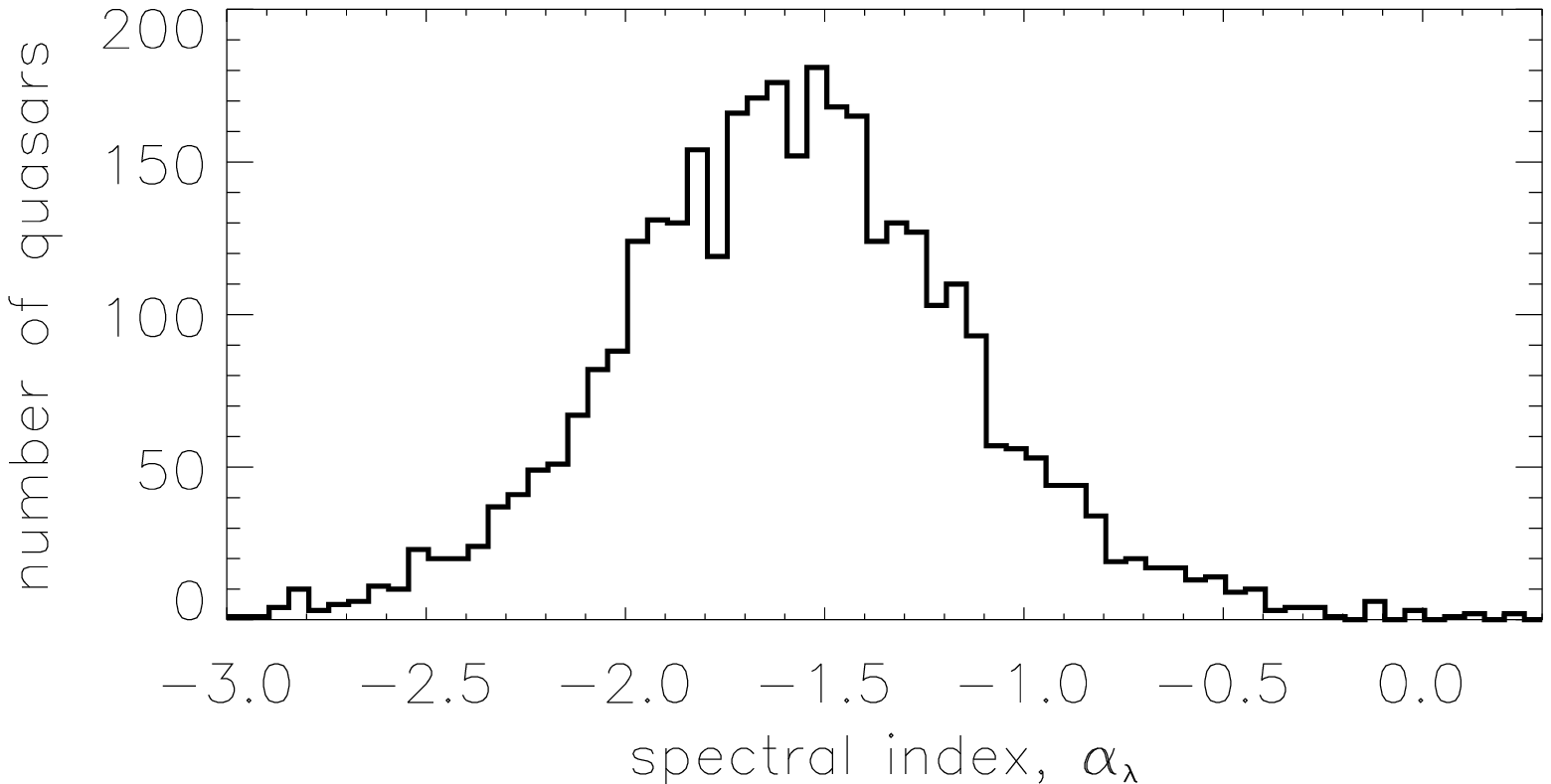}
 \end{tabular}
 \end{center}
 \vspace*{-4ex}
 \caption{Redshift distribution and spectral index distribtion of the sample of quasars}\label{fig:distr}
 \end{figure}

\begin{figure}[!h]
 \begin{center}
 \begin{tabular}{c}
   \epsfxsize=157mm
 \epsfbox{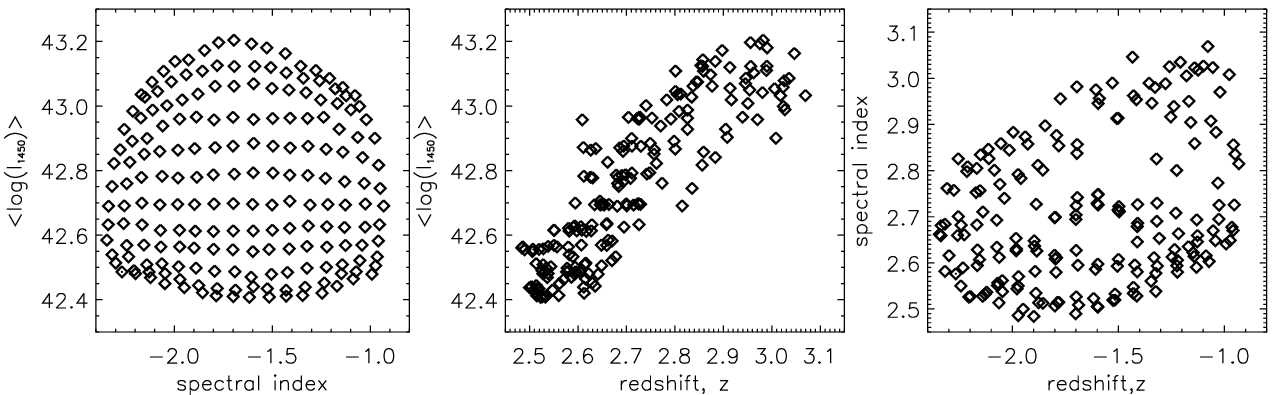}
 \end{tabular}
 \end{center}
 \vspace*{-4ex}
 \caption{Dependence between $\langle z \rangle$, $\langle$log $l_{1450} \rangle$ of subsamples and $\alpha_{\lambda}$ of composite spectra.}\label{fig:af}
 \end{figure}

\begin{figure}[!h]
\begin{center}
\begin{tabular}{c}
  \epsfxsize=75mm
\epsfbox{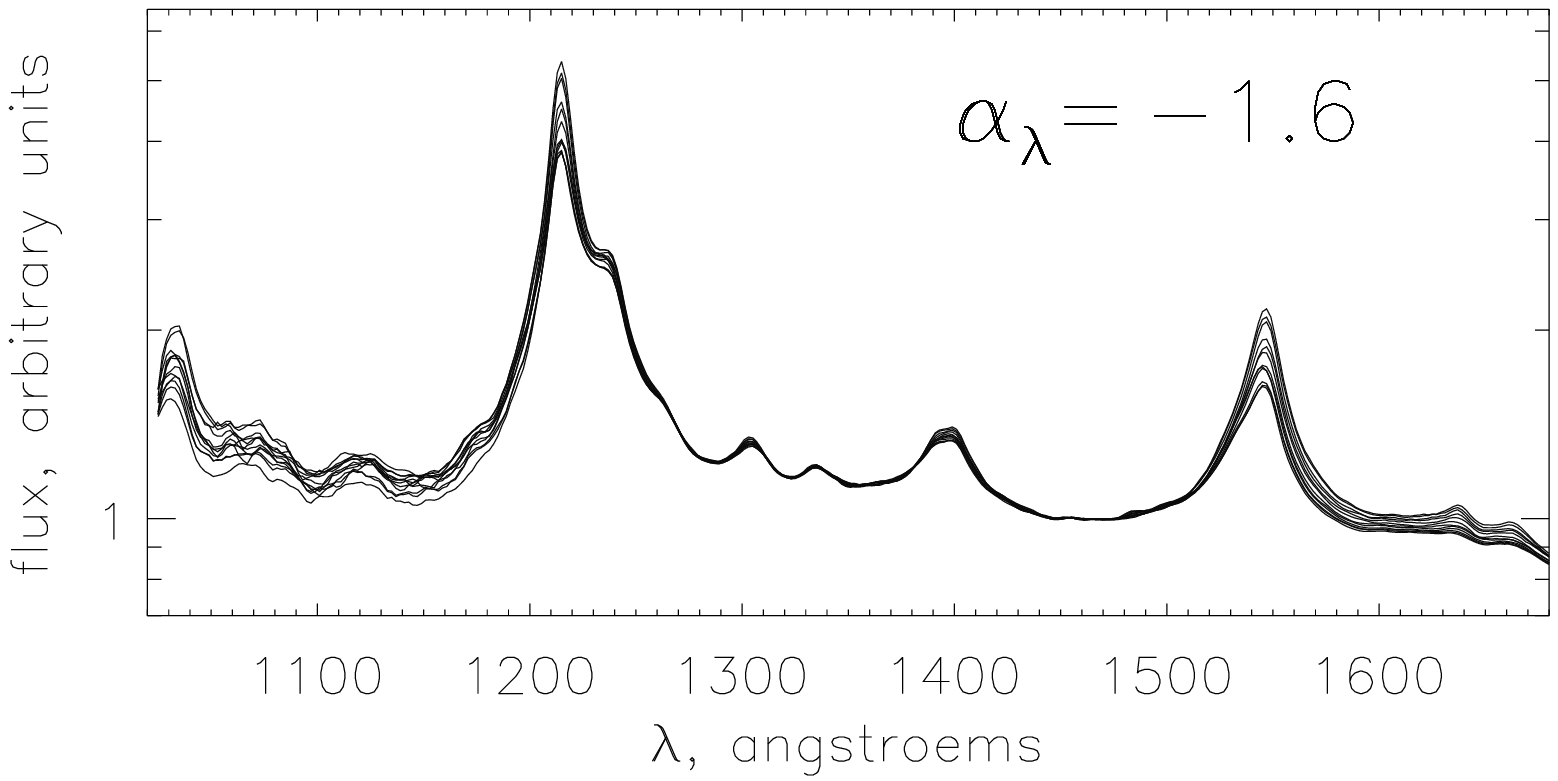}
  \epsfxsize=75mm
\epsfbox{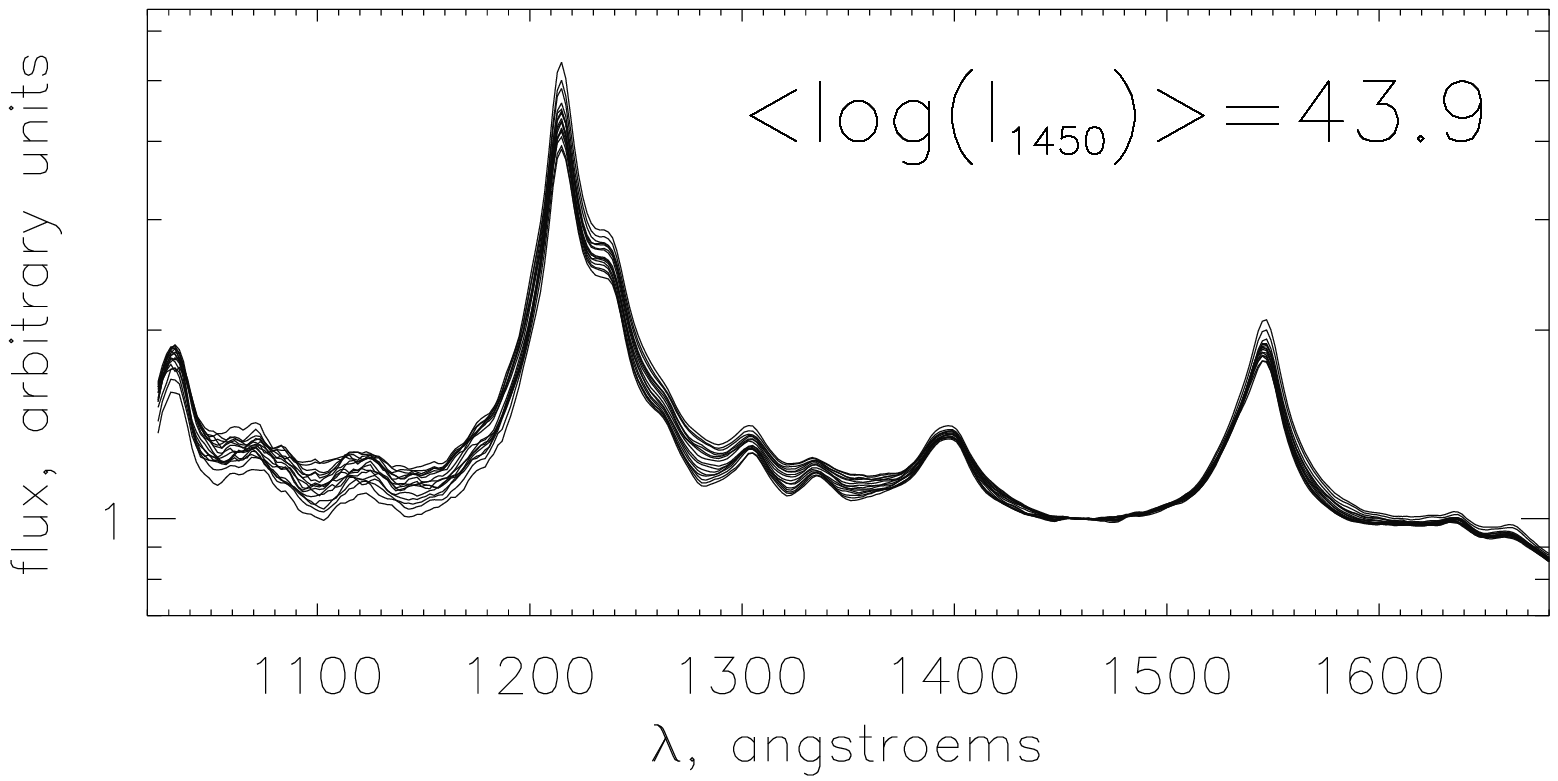}
\end{tabular}
\end{center}
\vspace*{-4ex}
 \caption{Subsets of composite spectra stacked from subsamples with the same $\alpha_{\lambda}$ but different $\langle\log{l_{1450}}\rangle$ (left, here the faintest ones are on top) and with the same $\langle\log{l_{1450}}\rangle$ but different $\alpha_{\lambda}$ (right, here the steepest are on top). }\label{fig:3-lum}
\end{figure}

We used the sample of 3535 individual quasar spectra from the Sloan Digital Sky Survey (SDSS) Data Release 7 with medium resolution of $R\sim2000$, compiled by \cite{ivashchenko+13}. Figure~\ref{fig:distr} shows the redshift and the spectral index distribution of this sample. We selected $16\times12=192$ subsamples with different spectral index $\alpha_{\lambda}$ (16 steps, from $-2.3$ to $-0.7$) and monochromatic luminosity at 1450\,\AA, $\log{l_{1450}}$ (12 steps, from 42.4 to 43.4), and then compiled 192 composite spectra from them. Here $l_{1450}$ is in erg\,s$^{-1}$\,\AA$^{-1}$. Figure~\ref{fig:af} shows the mean redshift, average logarithm of monochromatic luminosities at 1450\,\AA\ of subsamples and spectral index of composite spectra diagrams. Figure~\ref{fig:3-lum} shows examples of composite spectra bunches with the same spectral index and different luminosity and vice versa, correspondingly. 

\section{The method}

\begin{figure}[!h]
\begin{center}
\begin{tabular}{c}
\epsfxsize=155mm
\epsfbox{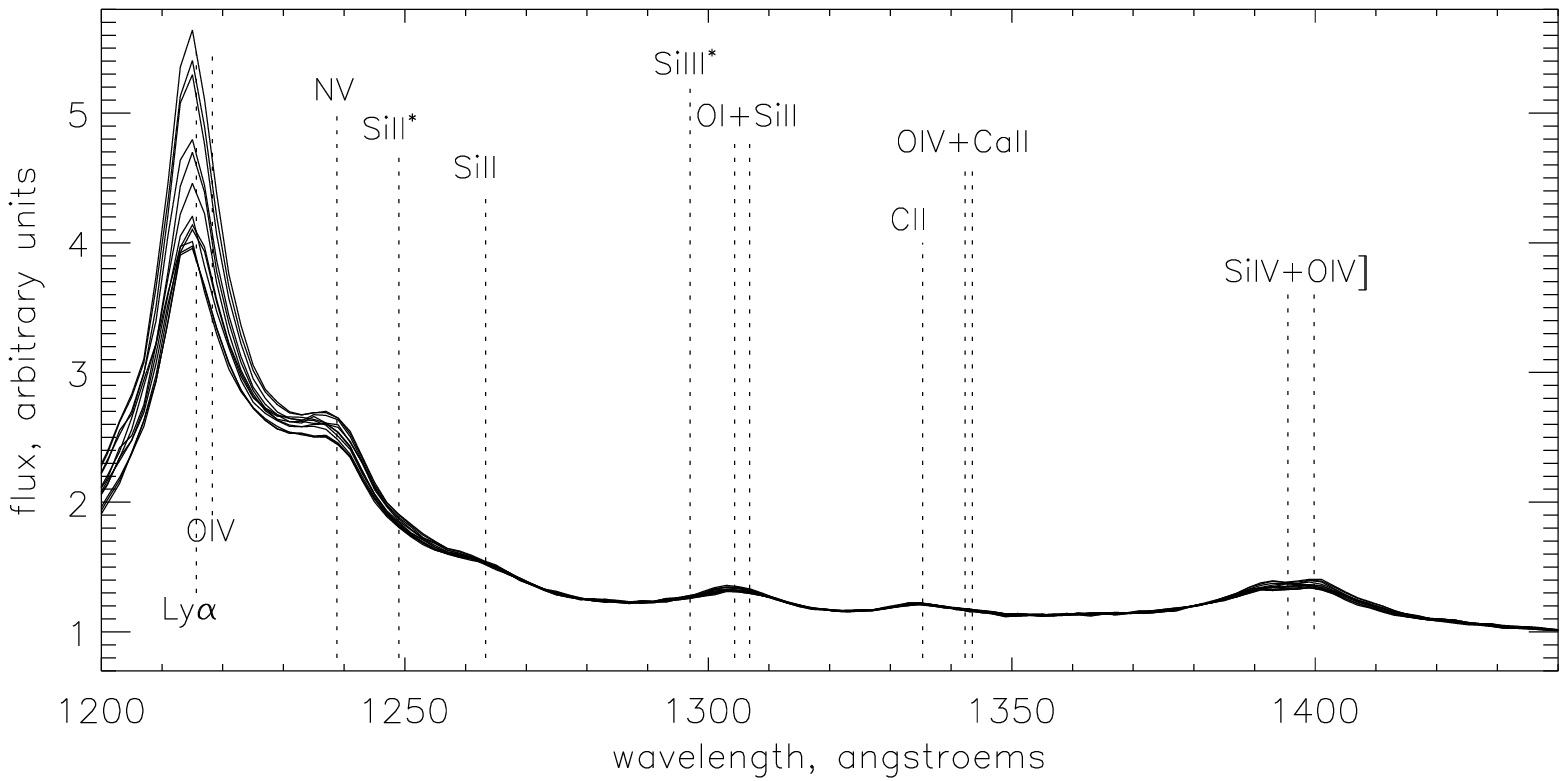}
\end{tabular}
\end{center}
\vspace*{-4ex}
\caption{Composite spectra of 12 subsamples with different luminosity and $\alpha_{\lambda}=-2.2$ with lines identified by \cite{vandenberk+01}. }\label{fig:comp}
\end{figure}

The wavelength ranges 1215--1285~\AA, 1290--1320~\AA, 1320--1350~\AA\ and 1350--1430~\AA\ were considered separately and fitted with the help of the \texttt{IDL lmfit} subroutine with a superposition of constant ($b=\mathrm{const}$) or power-law ($b=c\cdot\lambda^{\alpha_{\lambda}}$) continuum and the smallest possible number of Gaussian-profile emission lines in the following form:
\begin{equation}\label{eq:compos-1}
 f(\lambda)=b+\sum\limits_{k}a_{k}\exp\left[-\frac{(\lambda-\lambda^{0}_{k})^{2}}{2w_{k}^{2}}\right],
\end{equation}
where $\lambda$ is the rest-frame wavelength, $a_{k}$, $\lambda_{k}$, $w_{k}$ are the amplitude, the central wavelength and the full width at half maximum (FWHM) (up to $\sqrt{2}$-factor) of the $k$-th line. Then the equivalent widths of these four separate sets of lines were calculated using obtained model parameters. Figure~\ref{fig:comp} shows the whole wavelength range considered in the present study along with the known lines.

The virial mass of central supermassive black hole for each of 3535 individual quasar spectra was calculated using C~{\sc iv} emission line and the following empirical formula from \cite{shen+11}:
	\begin{equation}
           \log\left(\frac{M_{BH}}{M_{\odot}}\right) = a + b\log\left(\frac{\lambda L_{\lambda}}{10^{44} \mathrm{erg\,s}^{-1}}\right) + 2\log\left(\frac{W}{\mathrm{km\,s}^{-1}} \right),
       \end{equation}
where
       \begin{equation}
L_{\lambda} = 4\pi D^{2}_{phot}F_{\lambda},\quad  D_{phot} = \frac{c(1+z)}{H_{0}}\int\limits_{0}^{z}\frac{dt}{\sqrt{\Omega_{\Lambda} + \Omega_{M}(1+t)^{3}}}.
       \end{equation}
Here $F_{\lambda}$ and $L_{\lambda}$ are flux and luminosity, $W$ is the FWHM of C~{\sc iv} (1549~\AA), $D_{phot}$ is the photometric distance. We used the following values of cosmological parameters: $H_{0} = 67.79\pm0.78$ km\,s$^{-1}$\,Mpc$^{-1}$, $\Omega_{\Lambda} = 0.692\pm0.010$, $\Omega_{M} = 0.308\pm0.010$ from Planck Collaboration results (\cite{planck+13}), and calibration parameters $a = 0.66$ and $b = 0.53$  for C~{\sc iv} from \cite{shen+11}.

\section{Results and discussion}

\begin{figure}[!h]
 \begin{center}
 \begin{tabular}{c}
   \epsfxsize=78mm
    \epsfbox{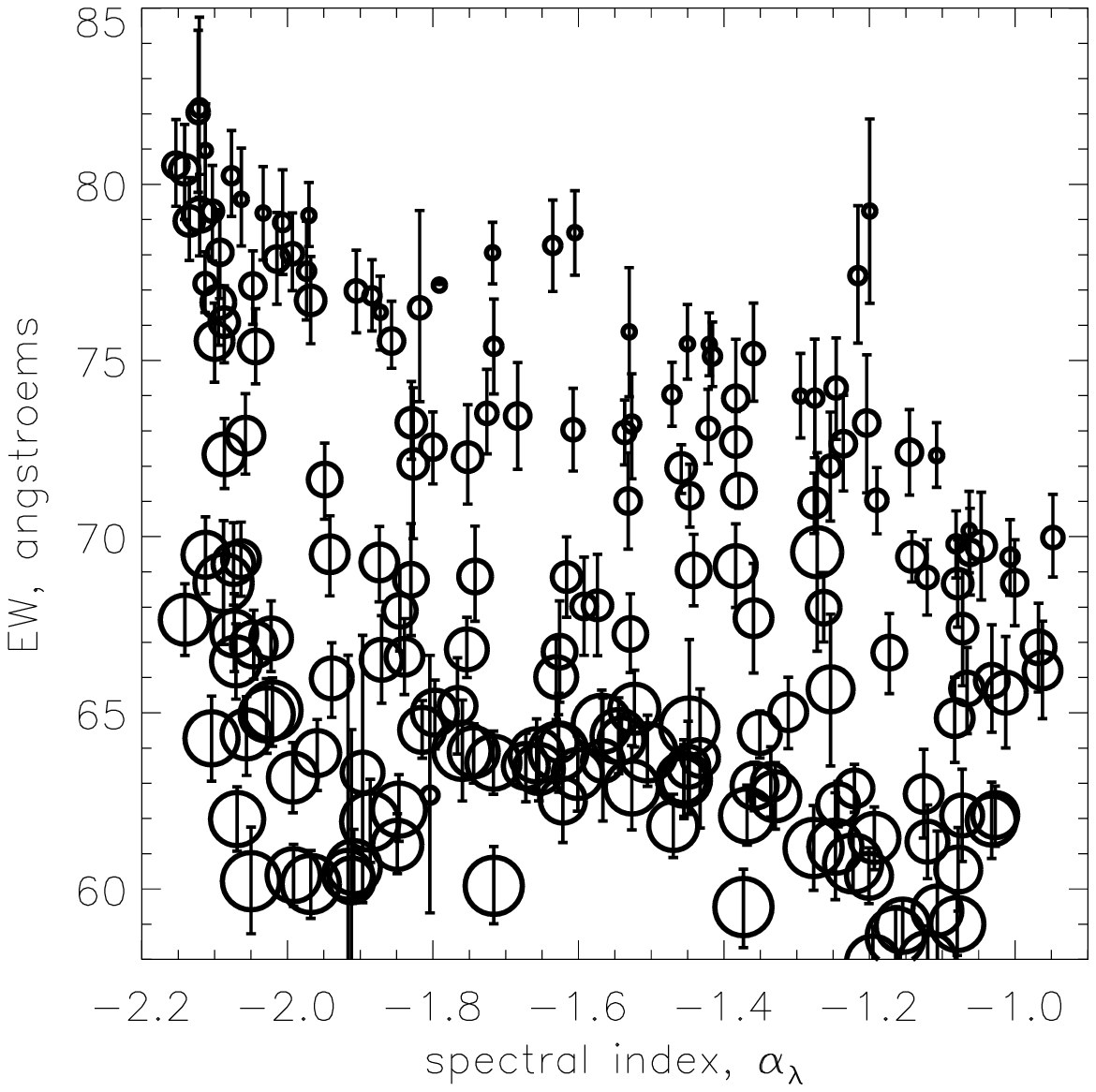}
   \epsfxsize=78mm
    \epsfbox{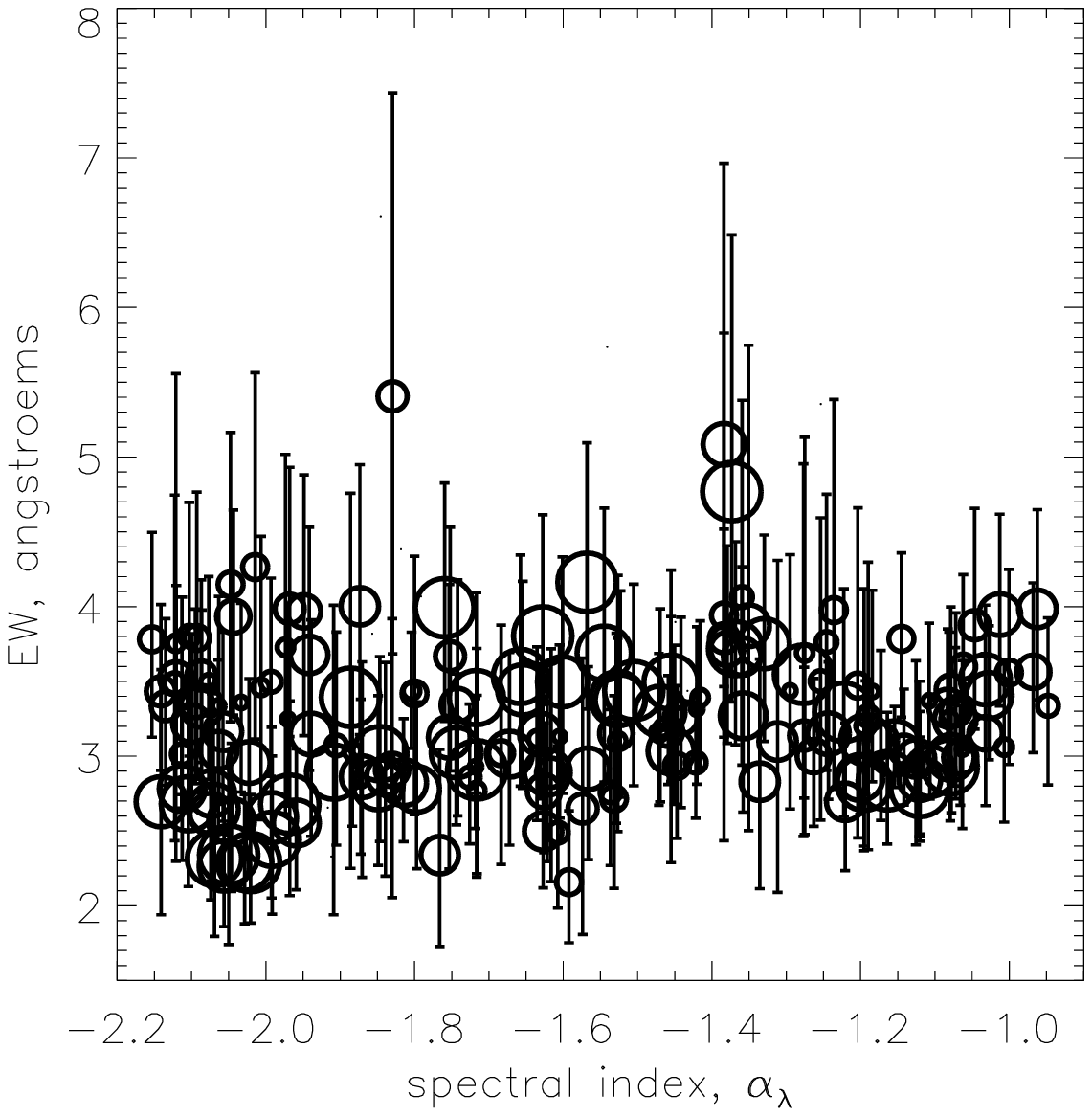}\\
   \epsfxsize=78mm
     \epsfbox{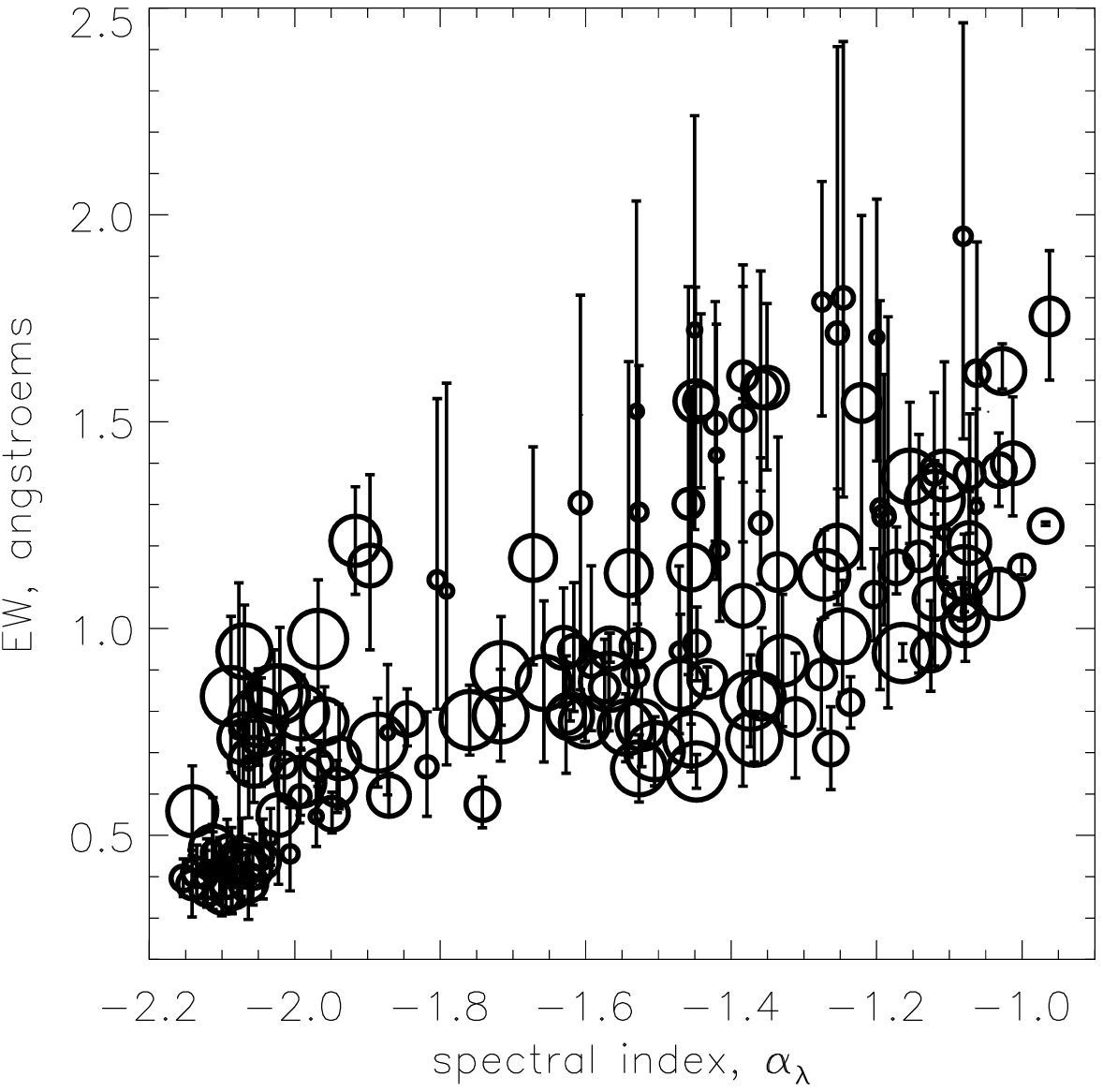}
   \epsfxsize=78mm
     \epsfbox{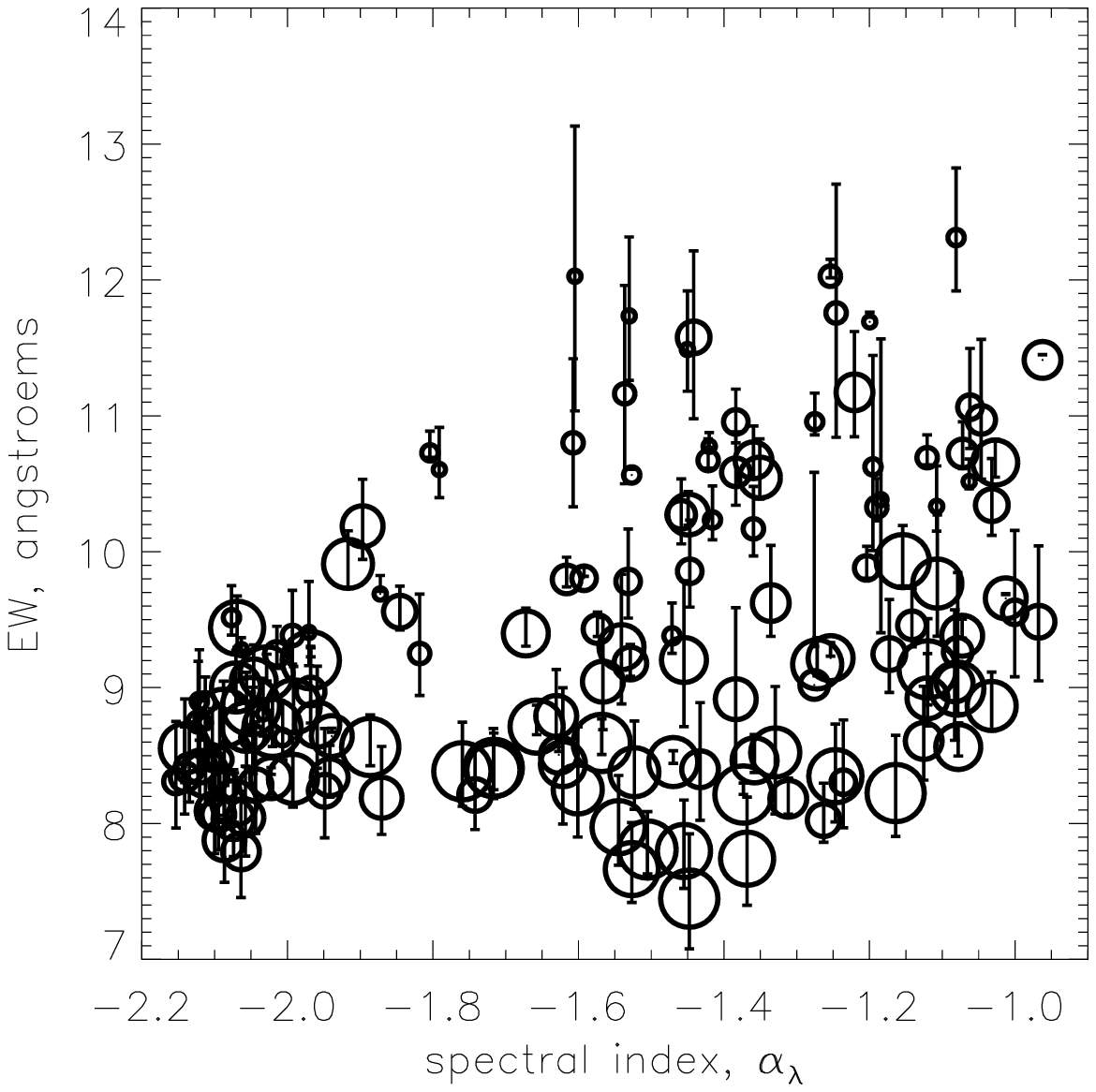}
 \end{tabular}
 \end{center}
 \vspace*{-4ex}
 \caption{Dependence of equivalent width of quasar emission lines on UV spectral index for superposition of lines within the wavelength ranges 1215--1285~\AA, 1290--1320~\AA, 1320--1350~\AA, 1350--1430~\AA\ (circle size shows the change of the luminosity). Here X1 and X2 are two bumps in this set of lines which were not identified. }\label{fig:res1}
 \end{figure}

The obtained $\alpha_{\lambda}$--EW diagrams for different luminosities are shown in Figure~\ref{fig:res1}. One can clearly see dependence between these two characteristics in some cases. In all cases the significance of possible dependence was checked with the F-test using constant and linear models as restricted and unrestricted ones, respectively. The resulting coefficients of the fitted models (the constant $b$ for the first case and $a_{1}$, $a_{2}$ for the second one) along with their errors, and the values of cumulative probability $P$ are presented in Tables~1, 2. Summarising the obtained results gives us the following:

(i) For lines Ly$\alpha$+O{\sc v}+N{\sc v}+Si{\sc ii$^{\ast}$}+Si{\sc ii} (1215--1285~\AA) the EW decreases with $\alpha_{\lambda}$; the significance of the linear coefficient of regression in most cases is $\sim3\sigma$, and the Baldwin effect is clearly seen for this set of lines. The more detailed analysis  with separation of this region into three parts showed the same trend for Ly$\alpha$+O{\sc v} and Si{\sc ii}, unlike N{\sc v}+Si{\sc ii$^{\ast}$} for which we did not found any dependence. At the same time, the Baldwin effect is seen for Ly$\alpha$+O{\sc v} and Si{\sc ii} lines, while for N{\sc v}+Si{\sc ii$^{\ast}$} we do not see any correlation between EW and luminosity, which agree with previous studies. Presence of the Baldwin effect was found by other authors for our first two parts together (Ly$\alpha$+O{\sc v}+N{\sc v}+Si{\sc ii$^{\ast}$}, see e.\,g. \cite{green+01,shang+03}). Some authors, e.\,g. \cite{bachev+04,shen+11,tang+12} tried to separate this set into two parts. All of them report on the Baldwin effect for Ly$\alpha$+O{\sc v}, and only \cite{shen+11} claims for Baldwin effect for N{\sc v}+Si{\sc ii$^{\ast}$} unlike two others who did not found Baldwin effect for this spectral feature.

(ii)  For lines Si{\sc iii$^{\ast}$}+O{\sc i}+Si{\sc ii} (1290--1320~\AA) the significance of the linear coefficient of regression differs for subsamples with different luminosity, moreover the values of $a_{2}$ differ in sign and have errors $50-100\%$, thus we cannot speak about some general trend. This confirms the absence of any dependence of EW on $\alpha_{\lambda}$ seen by eye. We also did not see any Baldwin effect for this set of lines, although it was found by \cite{shang+03} for O{\sc i} line.

(iii) Unlike the first case, for two other sets of lines, C{\sc ii}+O{\sc v}+Ca{\sc ii} (1290--1350~\AA) and  X1+Si{\sc iv}+O{\sc iv}]+X2 (1290--1350~\AA), we found the increase of EW with $\alpha_{\lambda}$, but only for the second set we can clearly see the Baldwin effect, found previously by other authors. On the other hand, note that the significance of the $\alpha_{\lambda}$--EW dependence in most cases are $\sim3\sigma$, but in the second case for some subsamples we have negative $a_{2}$ values and also large errors.

\begin{table}[!h]
\centering
 \caption{Coefficients of the fitted models: the constant $b$ for the constant fit and $a_{1}$, $a_{2}$ for the linear ($a_{1}+a_{2}\alpha_{\lambda}$), and the values of cumulative probability $P$ for F-test.}\label{tab:ftest-1}
% \vspace*{1ex}
\fontsize{9}{12}\selectfont
\begin{tabular}{|c|c|c|c|c|c|c|c|c|}
\hline
\multirow{2}{*}{$\log\langle{l_{1450}}\rangle$} & \multicolumn{4}{|c|}{Ly$\alpha$+O~{\sc v}+N~{\sc v}+Si~{\sc ii$^{\ast}$}+Si~{\sc ii}}  & \multicolumn{4}{|c|}{Si~{\sc iii$^{\ast}$}+O~{\sc v}+Si~{\sc ii}} \\
\cline{2-9}
  & $a_{1}$ & $a_{2}$ & $b$ & P & $a_{1}$ & $a_{2}$ & $b$ & P \\
\hline 
43.4 & 64.95$\pm$1.39 & -6.81$\pm$0.77 & 77.13$\pm$0.02 & $0.994$ & 3.35$\pm$0.74 &  0.10$\pm$0.05 & 3.19$\pm$0.16 & $0.210$  \\
43.5 & 61.72$\pm$1.48 & -8.52$\pm$0.94 & 74.85$\pm$0.33 & $0.994$ & 2.95$\pm$0.80 & -0.18$\pm$0.09 & 3.23$\pm$0.16 & $0.970$  \\
43.6 & 62.20$\pm$1.27 & -7.52$\pm$0.79 & 74.04$\pm$0.30 & $0.999$ & 3.09$\pm$0.67 &  0.02$\pm$0.01 & 3.06$\pm$0.15 & $0.960$ \\ 
43.7 & 61.24$\pm$1.36 & -7.63$\pm$0.82 & 73.57$\pm$0.30 & $0.998$ & 3.29$\pm$0.80 &  0.04$\pm$0.02 & 3.23$\pm$0.18 & $0.980$  \\
43.8 & 60.04$\pm$1.20 & -7.96$\pm$0.76 & 72.17$\pm$0.27 & $0.999$ & 3.01$\pm$0.60 & -0.21$\pm$0.09 & 3.32$\pm$0.15 & $0.880$ \\
43.9 & 59.03$\pm$1.19 & -7.41$\pm$0.76 & 70.35$\pm$0.26 & $0.999$ & 2.76$\pm$0.56 & -0.32$\pm$0.07 & 3.21$\pm$0.14 & $0.980$ \\
44.0 & 53.86$\pm$1.20 & -8.15$\pm$0.75 & 66.57$\pm$0.28 & $0.994$ & 3.37$\pm$0.66 &  0.20$\pm$0.10 & 3.07$\pm$0.16 & $0.150$  \\
44.1 & 55.10$\pm$1.21 & -6.67$\pm$0.76 & 65.47$\pm$0.26 & $0.993$ & 3.56$\pm$0.64 &  0.29$\pm$0.19 & 3.12$\pm$0.16 & $0.098$ \\
44.2 & 55.04$\pm$1.16 & -5.57$\pm$0.72 & 63.83$\pm$0.24 & $0.990$ & 3.47$\pm$0.52 &  0.39$\pm$0.33 & 2.88$\pm$0.13 & $0.999$  \\
44.3 & 57.68$\pm$1.35 & -3.46$\pm$0.80 & 63.38$\pm$0.27 & $0.830$ & 3.56$\pm$0.46 &  0.42$\pm$0.31 & 2.97$\pm$0.12 & $0.999$  \\
44.4 & 56.12$\pm$1.16 & -3.48$\pm$0.71 & 61.69$\pm$0.25 & $0.960$ & 3.73$\pm$0.54 &  0.53$\pm$0.35 & 2.93$\pm$0.14 & $0.998$  \\
44.5 & 54.77$\pm$1.34 & -4.43$\pm$0.85 & 61.65$\pm$0.27 & $0.990$ & 3.45$\pm$0.69 &  0.34$\pm$0.04 & 2.95$\pm$0.17 & $0.994$ \\
\hline
\end{tabular}
\end{table}

\begin{table}[!h]
\centering
 \caption{Coefficients of the fitted models: the constant $b$ for the constant fit and $a_{1}$, $a_{2}$ for the linear fit ($a_{1}+a_{2}\alpha_{\lambda}$), and the values of cumulative probability $P$ for F-test.}\label{tab:ftest-2}
% \vspace*{1ex}
\fontsize{9}{11}\selectfont
\begin{tabular}{|c|c|c|c|c|c|c|c|c|}
\hline
\multirow{2}{*}{$\log\langle{l_{1450}}\rangle$} & \multicolumn{4}{|c|}{C~{\sc ii}+O~{\sc v}+Ca~{\sc ii}} & \multicolumn{4}{|c|}{X1+Si~{\sc iv}+O~{\sc iv}]+X2} \\
\cline{2-9}
  & $a_{1}$ & $a_{2}$ & $b$ & P & $a_{1}$ & $a_{2}$ & $b$ & P \\
\hline 
43.4 & 2.31$\pm$0.20 & 0.88$\pm$0.11 & 0.68$\pm$0.04 & $0.999$ & 14.92$\pm$0.07 &  2.76$\pm$0.05 & 11.15$\pm$0.01 & $0.999$ \\
43.5 & 2.95$\pm$0.33 & 1.20$\pm$0.17 & 0.63$\pm$0.05 & $0.999$ & 12.98$\pm$0.28 &  1.58$\pm$0.18 & 10.56$\pm$0.01 & $0.999$ \\
43.6 & 2.71$\pm$0.25 & 1.08$\pm$0.13 & 0.56$\pm$0.03 & $0.999$ & 14.76$\pm$0.12 &  2.78$\pm$0.06 &  9.08$\pm$0.01 & $0.999$ \\
43.7 & 1.77$\pm$0.05 & 0.62$\pm$0.03 & 0.92$\pm$0.02 & $0.999$ & 11.69$\pm$0.29 &  1.21$\pm$0.18 &  9.79$\pm$0.03 & $0.997$ \\
43.8 & 1.81$\pm$0.12 & 0.63$\pm$0.06 & 0.70$\pm$0.03 & $0.999$ & 12.39$\pm$0.30 &  1.79$\pm$0.19 &  9.62$\pm$0.05 & $0.999$ \\
43.9 & 1.95$\pm$0.02 & 0.72$\pm$0.02 & 1.22$\pm$0.01 & $0.999$ & 10.14$\pm$0.38 &  0.68$\pm$0.23 &  9.05$\pm$0.07 & $0.970$ \\
44.0 & 1.90$\pm$0.09 & 0.72$\pm$0.05 & 0.79$\pm$0.02 & $0.999$ & 15.43$\pm$0.06 &  4.19$\pm$0.07 & 11.41$\pm$0.01 & $0.999$ \\
44.1 & 2.42$\pm$0.14 & 0.95$\pm$0.08 & 0.78$\pm$0.02 & $0.999$ & 10.82$\pm$0.02 &  1.15$\pm$0.01 &  9.58$\pm$0.01 & $0.999$ \\
44.2 & 2.34$\pm$0.11 & 0.95$\pm$0.07 & 0.83$\pm$0.01 & $0.999$ & 10.26$\pm$0.18 &  0.83$\pm$0.13 &  9.14$\pm$0.04 & $0.990$ \\
44.3 & 1.49$\pm$0.12 & 0.40$\pm$0.07 & 0.84$\pm$0.03 & $0.999$ &  8.33$\pm$0.28 & -0.18$\pm$0.08 &  8.60$\pm$0.04 & $0.860$ \\ 
44.4 & 1.02$\pm$0.15 & 0.13$\pm$0.09 & 0.82$\pm$0.03 & $0.999$ &  8.58$\pm$0.25 & -0.18$\pm$0.12 &  8.88$\pm$0.05 & $0.870$ \\
44.5 & 1.28$\pm$0.09 & 0.30$\pm$0.06 & 0.88$\pm$0.02 & $0.880$ &  6.32$\pm$0.14 & -1.40$\pm$0.08 &  8.79$\pm$0.02 & $0.998$ \\
\hline
\end{tabular}
\end{table}
\vspace*{1ex}

\begin{figure}[!h]
 \begin{center}
 \begin{tabular}{c}
   \epsfxsize=50mm
   \epsfbox{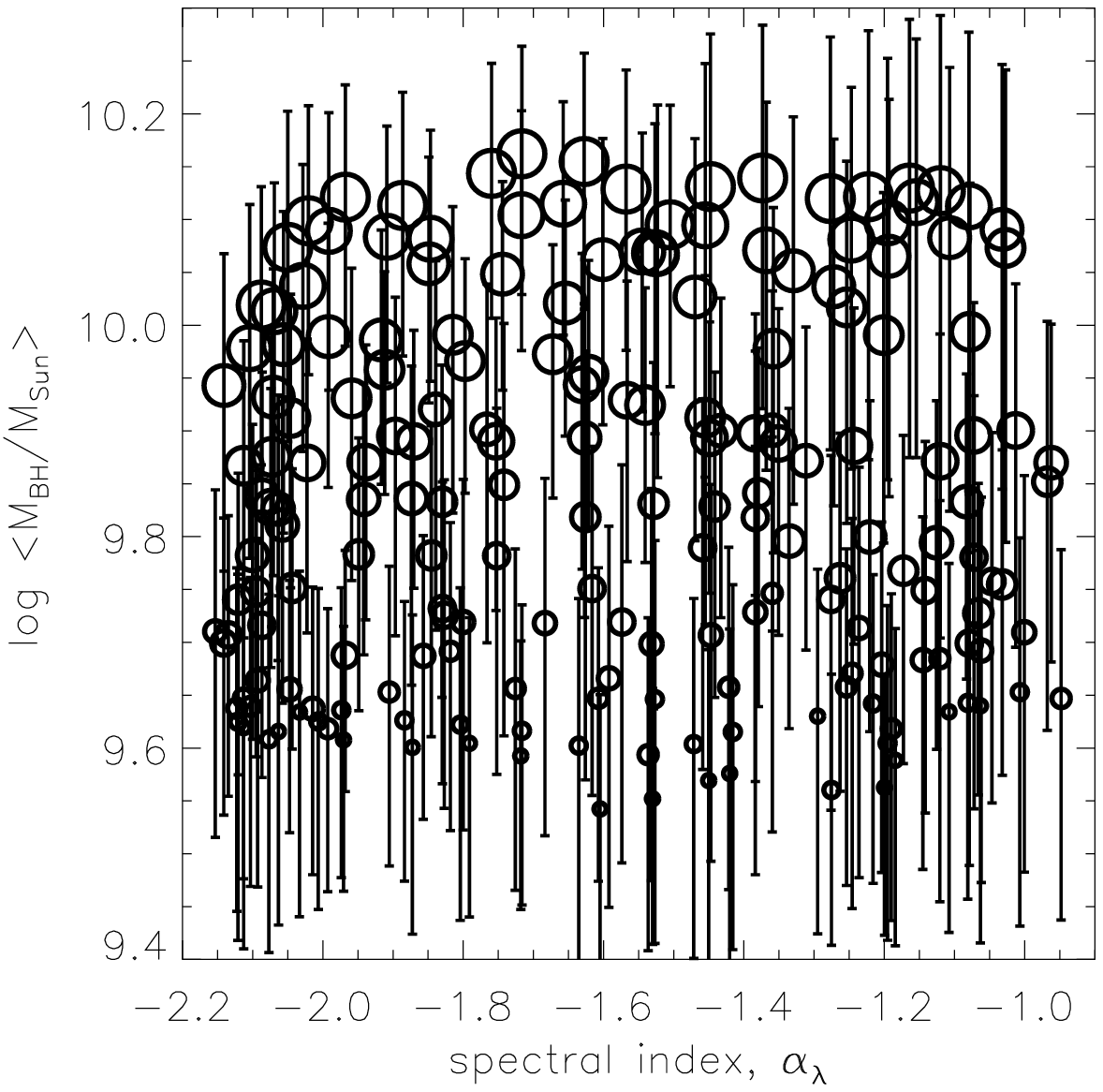}
   \epsfxsize=50mm
   \epsfbox{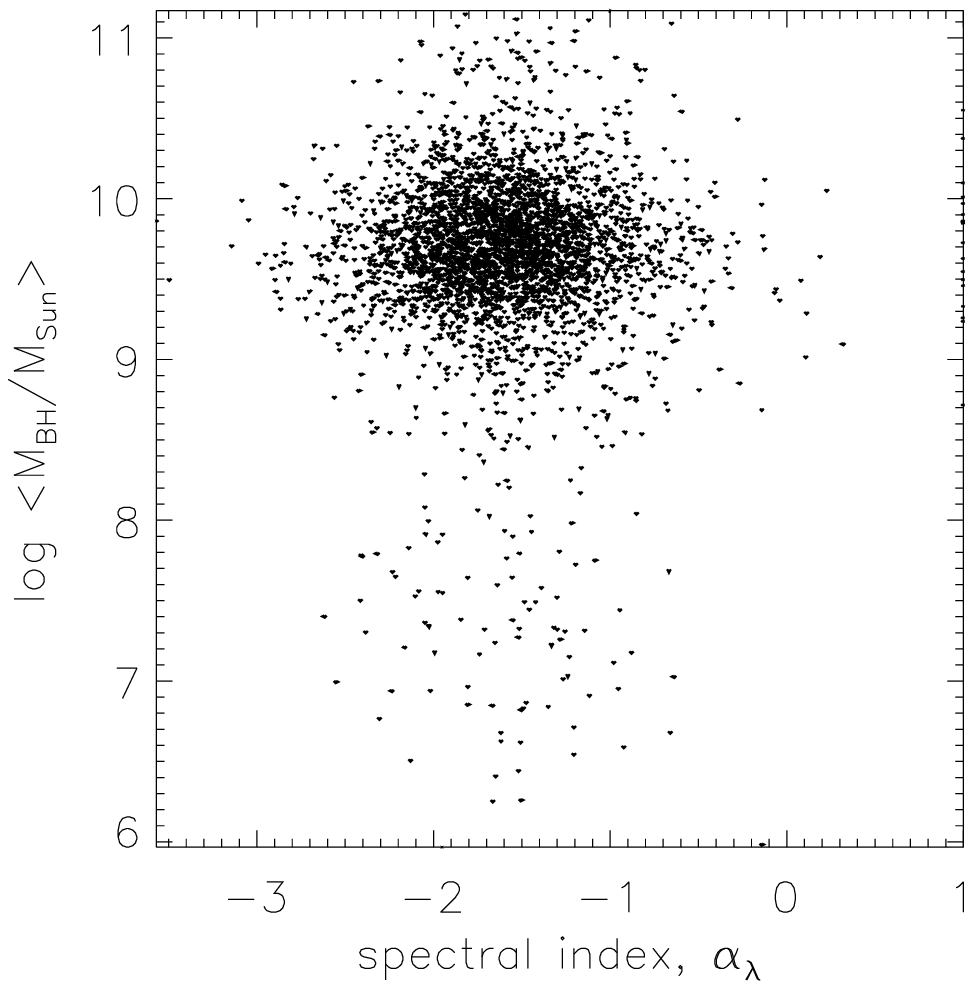}
      \epsfxsize=50mm
   \epsfbox{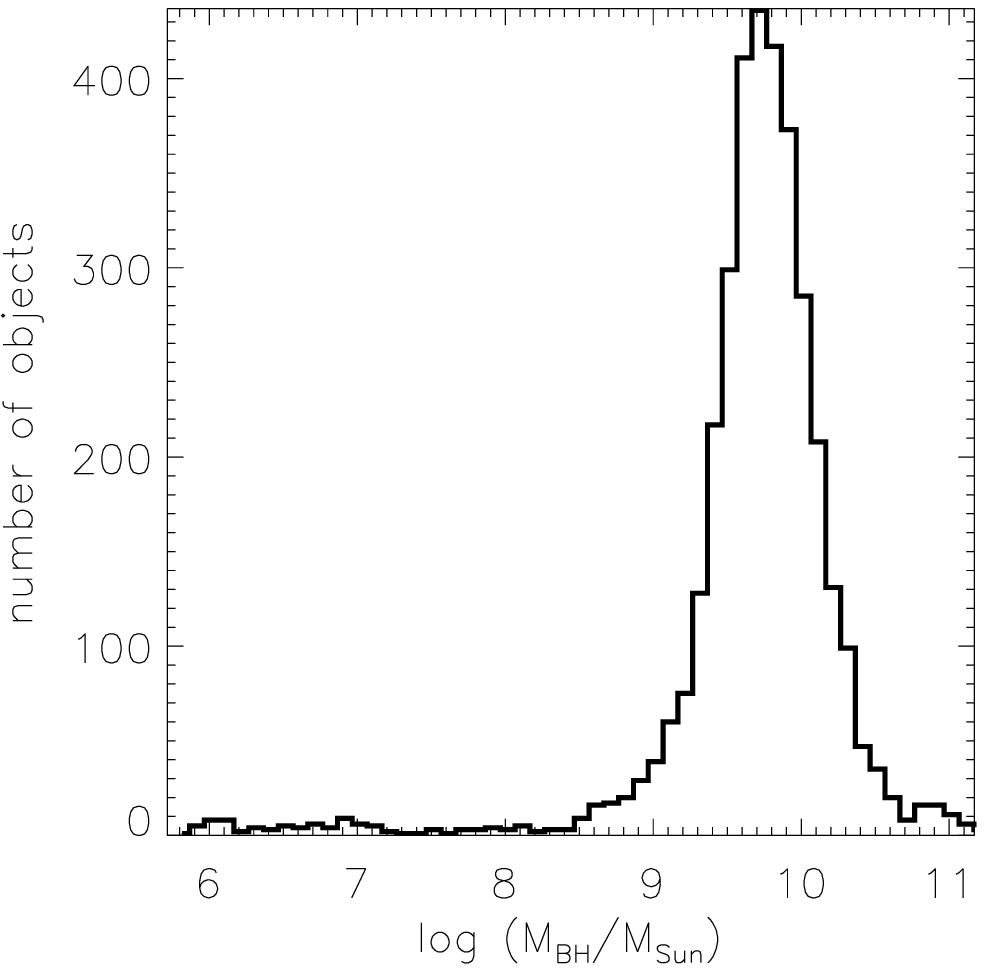}
 \end{tabular}
 \end{center}
 \vspace*{-4ex}
 \caption{Mass of quasar's BH -- $\alpha_{\lambda}$ diagram for composite spectra (left; circle size shows luminosity changes) and for 3535 individual quasar (centre), and BH mass distribution (right).}\label{fig:mass}
 \end{figure}

Figure~\ref{fig:mass} shows $\alpha_{\lambda}$--M$_{BH}$ diagrams for composite spectra (with different luminosities) on the left and for all 3553 individual quasars on the right. It is clearly seen that there is no dependence in both cases.

Unlike the case of relation between the monochromatic luminosity of the quasar and the EW of its emission lines (the Baldwin effect), interpretation of relation between the spectral index, $\alpha_{\lambda}$, and the EW is not so simple. Firstly, one have to remember that the term `spectral index' describes here only a slope of some limited part of the UV-optical quasar SED (the Big Blue Bump). It has nothing common with a real power-law spectra inherent to non-thermal radiation, and serves here only as tool for some classification of quasar SEDs. The Big Blue Bump is considered to be a result of thermal emission of inhomogeneous gas-dust torus, i.\,e. a superposition of a number of Planck curves with different temperature. It has been shown previously that there is no correlation between $\alpha_{\lambda}$ and $\log{l_{1450}}$ (\cite{ivashchenko+13}) and in the present work we also show that there is no dependence between $\alpha_{\lambda}$ and the BH mass. Hence we can assume that the shape of UV-optical SED of quasars does not depend, at least directly, on the central BH. It has to be influenced by the physical properties of the torus and can also be affected by the host galaxy. On the other hand, the obtained dependence between $\alpha_{\lambda}$ and EW of some emission lines, mostly for those lines for which the Baldwin effect is observed, claims for some relation between the physical properties of the torus and the BLR, e.\,g. the chemical abundance.

\acknowledgments %\footnotesize
This work has been supported by the Complex Programme of Scientific Space Studies of the NAS of Ukraine for 2013-2016 and by the Swiss National Science Foundation grant SCOPE IZ7370-152581.

\end{article}
\end{document}